\documentclass[showpacs,showkeys,twocolumn,amsmath,amssymb,pra,superscriptaddress]{revtex4-1}

\usepackage{graphicx}

\newcommand{\ri}{{\rm i}}

\begin{document}

\author{Christoph Weiss}
\affiliation{Institut f\"ur Physik, Carl von Ossietzky Universit\"at, 26111 Oldenburg, Germany}
\affiliation{Department of Physics, Durham University, Durham DH1 3LE, United Kingdom}
\email{Christoph.Weiss@durham.ac.uk}

\title{Distinguishing mesoscopic quantum superpositions from statistical mixtures in periodically shaken double wells}

\date{22 October 2011}

\begin{abstract}
For Bose-Einstein condensates in double wells,  $N$-particle Rabi-like oscillations often seem to be damped. Far from being a decoherence effect, the apparent damping can indicate the emergence of quantum superpositions in the many-particle quantum dynamics. However, in an experiment it would be difficult to distinguish the apparent damping from decoherence effects. The present paper suggests using controlled periodic shaking to quasi-instantaneously switch the sign of an effective Hamiltonian, thus implementing an ``echo'' technique which distinguishes quantum superpositions from statistical mixtures. The scheme for the effective time-reversal is tested by numerically solving the time-dependent $N$-particle Schr\"odinger equation.

\end{abstract}




 \maketitle 

Small Bose-Einstein condensates (BECs) of some 1000~\cite{AlbiezEtAl05} or even 100 atoms~\cite{ChuuEtAl05} have been a topic of experimental research for several years. Recently, the investigation of many-particle wave-functions of BECs  in phase space became experimentally feasible~\cite{ZiboldEtAl10}. This experimental technique will further investigations of beyond-mean-field (Gross-Pitaevskii) behaviour for small BECs. 

\begin{figure}[h]
\includegraphics[width=\linewidth]{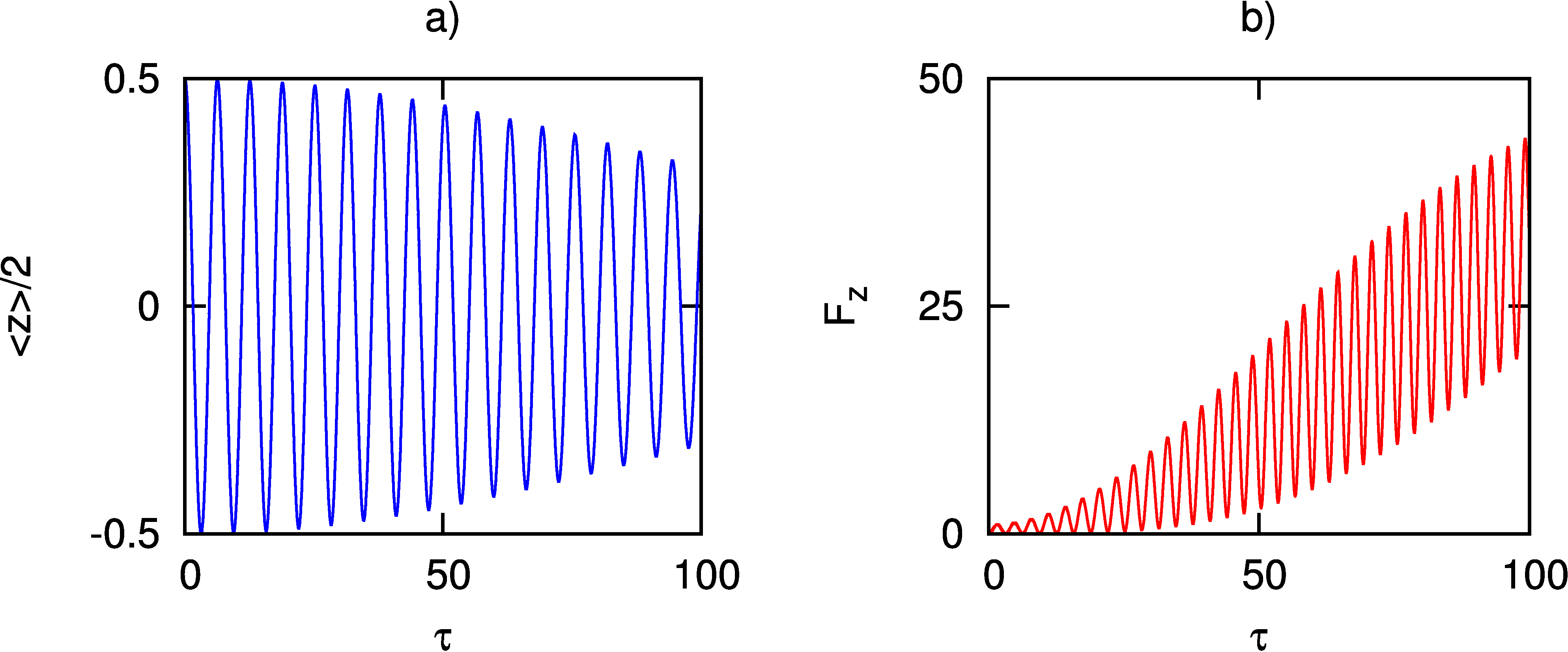}
\caption{\label{fig:damping}(Colour online) \textbf{a)} Population imbalance $\langle z\rangle/2$ [Eq.~(\ref{eq:popu})] as a function of dimensionless time $\tau$~(\ref{eq:tau}) for small BEC in a double-well potential. Initially, $N=100$ atoms are in well~1, the quantum dynamics is given by the Hamiltonian~(\ref{eq:H2mode}) ($NU/J=0.4$).  \textbf{b)} Variance of the population imbalance~(\ref{eq:Fz}). Experimentally, it will be difficult to distinguish the apparent damping and the increased fluctuations (which are both triggered by a collapse and revival phenomenon) from true damping introduced, e.g., by decoherence.}
\end{figure}

For a BEC initially loaded into one of the wells of a double-well potential, the many-particle oscillations often seem to be damped compared to the mean-field behaviour. Figure~\ref{fig:damping} shows such an apparent damping, which in fact is a collapse which will eventually be followed by at least partial revivals (cf.\  Refs.~\cite{HolthausStenholm01,Ziegler11}), for $N=100$ particles. This apparent damping coincides with an increase of the fluctuations of the number of particles in each well [Fig.~\ref{fig:damping}~(b)]. 

In order to numerically calculate the many-particle dynamics, the Hamiltonian in the two-mode approximation~\cite{MilburnEtAl97} is used,
\begin{eqnarray}
\label{eq:H2mode}
\hat{H}_{0} &=& 
-J \left(\hat{c}_1^{\dag}\hat{c}_{2}^{\phantom{\dag}}+\hat{c}_2^{{\dag}}\hat{c}_{1}^{\phantom{\dag}}\right)
+\frac U2 \sum_{j=1}^2\hat{n}_j\left(\hat{n}_j-1\right)\;,
\end{eqnarray}
where $\hat{c}_j^{(\dag)}$ are the boson creation and annihilation operators
on site $j$, $\hat{n}_j = \hat{c}_j^{\dag}\hat{c}_j^{\phantom{\dag}}$ are the number operators, $J$ is the hopping matrix element and $U$ the on-site interaction energy.

The experimentally measurable~\cite{EsteveEtAl08} population imbalance is useful to quantify the oscillations depicted in Fig.~\ref{fig:damping}:
\begin{equation}
\label{eq:popu}
  \frac{\langle z\rangle (\tau)}2 \equiv\frac{\langle n_2\rangle(\tau)-\langle n_1\rangle(\tau)}{2N}\;,
\end{equation}
where $\tau$ is the dimensionless time: 
\begin{equation}
\label{eq:tau}
\tau\equiv \frac{tJ}{\hbar}\;.
\end{equation}
The variance of the population imbalance can be quantified by using the experimentally measurable~\cite{EsteveEtAl08} quantity
\begin{equation}
\label{eq:Fz}
F_z \equiv \frac{\langle \Delta n_{12}^2\rangle}{N}\;,
\end{equation}
with $0\le F_z\le N$. For pure states, Eq.~(\ref{eq:Fz}) coincides with a quantum Fisher information~\cite{PezzeSmerzi09}. Like the spin-squeezed states investigated in Ref.~\cite{EsteveEtAl08} (and references therein), quantum superpositions with large fluctuations are also relevant to improve interferometric measurements beyond single-particle limits. A prominent example of a quantum superposition relevant for interferometry are the NOON-states~\cite{GiovannettiEtAl04}
\begin{equation}
\left|\psi_{\rm NOON}\right> = \frac1{\sqrt{2}}\left(\left|N,0\right>+\left|0,N\right>\right)\;,
\end{equation}
i.e., quantum superpositions of all particles either being in well one or in well two; $\left|n_1,n_2\right>$ refers to the Fock state with $n_1$ particles in well~1 and  $n_2$ particles in well~2. Suggestions how such states can be obtained for ultra-cold atoms can be found in Refs.~\cite{Ziegler11,MicheliEtAl03,MahmudEtAl05,StreltsovEtAl09,DagninoEtAl09,GertjerenkenEtAl10,GarciaMarchEtAl11,MazzarellaEtAl11} and references therein.
For pure states, $F_z>1$ indicates that this quantum superposition is relevant for interferometry~\cite{PezzeSmerzi09}. However, it remains to be shown that the increased fluctuations are really due to pure states rather than statistical mixtures.

It might sound tempting to use the revivals investigated in Refs.~\cite{HolthausStenholm01,Ziegler11} to identify pure quantum states. However, while such revivals can be observed, e.g., for two-particle systems~\cite{Folling07}, the situation for a BEC in a double well is more complicated. In principle, very good revivals of the initial wave-function should occur as long as the system is described by the Hamiltonian~(\ref{eq:H2mode}). While partial revivals can easily be observed, (nearly) perfect revivals might occur for times well beyond experimental time-scales -- in particular if the experiment is performed under realistic conditions subject to decoherence effects\footnote{For computer simulations, numerical errors might produce an effective decoherence which would again prevent nearly perfect revivals from occurring at very long time-scales.}. It is thus not obvious how such an apparent damping might be distinguished experimentally from decoherence effects which would lead to statistical mixtures with (now truly) damped oscillation similar to Fig.~\ref{fig:damping}. The focus of this paper  thus lies on an experimentally realisable ``echo'' technique to distinguish statistical mixtures from quantum superpositions by using periodic shaking.

\begin{figure}
\includegraphics[width=\linewidth]{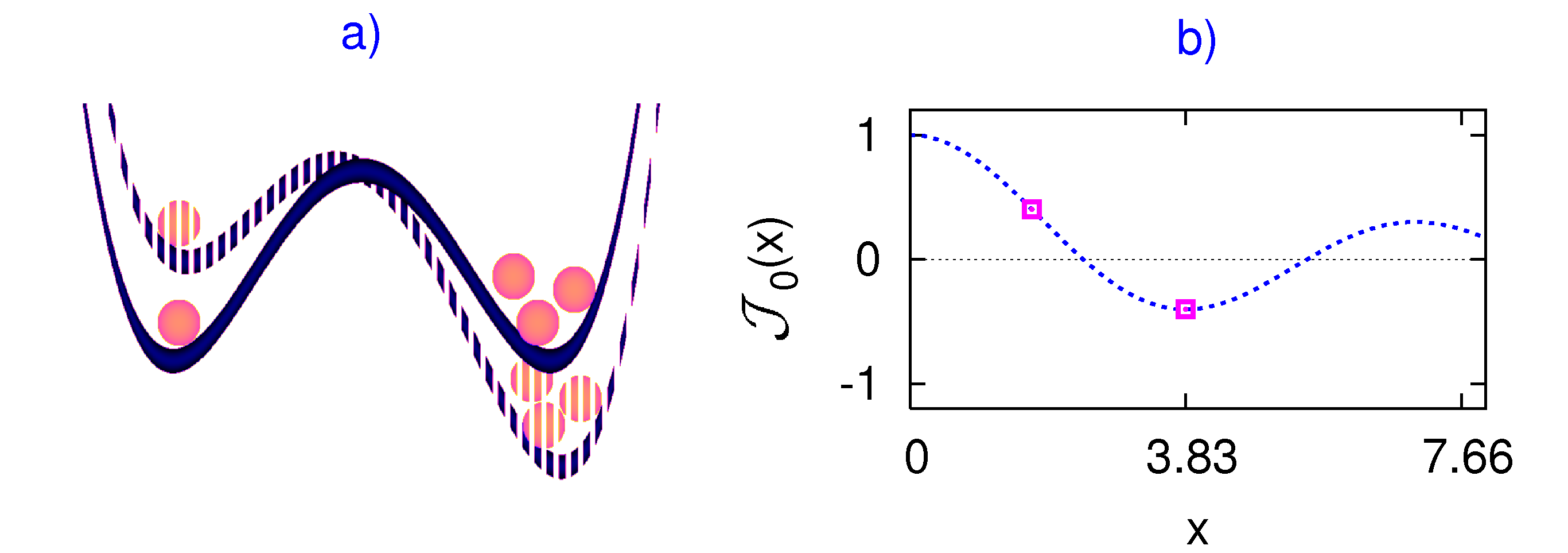}
\caption{\label{fig:sketch}(Colour online)  {\bf a)} Sketch of a double well which is shaken periodically to control tunnelling for ultra-cold atoms. 
{\bf b)} For high shaking frequencies, the tunnelling rate is modified by the ${\cal J}_0$-Bessel function [see Eq.~(\ref{eq:jeff})]. The two squares indicate a pair of shaking amplitudes for which the Bessel function has equal modulus and opposite sign ($x_1= 1.691732695$ and $x_2=3.831705970$ with $\left|J_0\left(x_{1,2}\right)\right|\simeq 0.403$ -- it will be shown in Fig.~\ref{fig:karten}~(d) that it is not essential to precisely use these values).}
\end{figure}

 Periodic shaking~\cite{GrifoniHanggi98} is currently being established experimentally to control tunnelling of BECs~\cite{SiasEtAl07,HallerEtAl10,Struck2011,ChenEtAl11,MaEtAl11,CiampiniEtAl2011}. For the model~(\ref{eq:H2mode}), periodic shaking can be included via
\begin{equation}
\label{eq:H2time}
\hat{H} = \hat{H}_{0} +\frac K2\cos(\omega t)(\hat{n}_2-\hat{n}_1)\;,
\end{equation}
where $K$ is the strength of shaking and $\omega$ its (angular) frequency. For large shaking frequencies\footnote{While the validity of this approximation also depends on the values chosen for the interaction, driving frequencies as low as $\hbar\omega \approx 6J$ can sometimes be considered large. Choosing higher frequencies will improve the approximation. However, as this will, in general, also increase the driving amplitude, for too high frequencies the two-mode approximation~(\ref{eq:H2mode}) no longer is valid.} and not-too-large interactions, the time-dependent Hamiltonian~(\ref{eq:H2time}) can be replaced by a time-independent effective Hamiltonian:
\begin{eqnarray}
\label{eq:Heff}
\hat{H}_{\rm eff} &=& 
-J_{\rm eff}\left(\hat{c}_1^{\dag}\hat{c}_{2}^{\phantom{\dag}}+\hat{c}_2^{{\dag}}\hat{c}_{1}^{\phantom{\dag}}\right)
+\frac U2 \sum_{j=1}^2\hat{n}_j\left(\hat{n}_j-1\right)
\end{eqnarray}
with
\begin{equation}
\label{eq:jeff}
J_{\rm eff} = J \mathcal{J}_0\left(K_0\right);\quad K_0\equiv{\frac{K}{\hbar\omega}}
\end{equation}
where $\mathcal{J}_0$ is the Bessel-function depicted in Fig.~\ref{fig:sketch}~(b).
Such effective Hamiltonians have  been successfully tested experimentally in optical lattices, see, e.g., Refs.~\cite{SiasEtAl07,CreffieldEtAl10}; negative $J_{\rm eff}$ have been experimentally investigated in Refs.~\cite{Struck2011,CiampiniEtAl2011}. There are, however, also examples~\cite{TeichmannEtAl09,EsmannEtAl11b} for which two or more Bessel function are needed to understand the tunnelling dynamics.

In the present situation, the effective description~(\ref{eq:Heff}) offers the possibility to quasi-instantaneously switch the sign of both the kinetic energy (via shaking, cf.~Fig.~\ref{fig:sketch}) and the interaction (via a Feshbach-resonance~\cite{BauerEtAl09}). Contrary to special cases where the wave-function~\cite{MorigiEtAl02,Meunier05} can be changed to obtain time-reversal, for periodically driven systems the \emph{Hamiltonian} can be changed by quasi-instantaneously changing both the tunnelling term [by switching the shaking amplitude, e.g., between values shown in Fig.~\ref{fig:sketch}~(b)] and the sign of the interaction via a Feshbach-resonance~\cite{BauerEtAl09};
\begin{equation}
\label{eq:hdream}
\hat{H}_{\rm ideal} \equiv \left\{  
\begin{array}{lcr}
  +\hat{H}_{\rm eff}(\tau\!=\!0)&:&\tau < \tau_0\\
  -\hat{H}_{\rm eff}(\tau\!=\!0)&:&\tau\ge \tau_0
  \end{array}\right..
\end{equation} 
The corresponding unitary time-evolution is given by 
\begin{equation}
U(0,\tau) = 
\left\{  
\begin{array}{lcr}
\exp\left(-\frac{\ri\tau \hat{H}_{\rm eff}(\tau=0)}{\hbar J}\right)&:&\tau < \tau_0\\
\exp\left(\frac{\ri(\tau-2\tau_0) \hat{H}_{\rm eff}(\tau=0)}{\hbar J}\right)&:&\tau\ge \tau_0
  \end{array}\right.,
\end{equation}
with perfect return to the initial state at $\tau=2\tau_0$.  However, the turning point $\tau_0$ has to be chosen with care: only by taking $\tau_0$ close to the maximum of the shaking can unwanted excitations be excluded  (cf.\ Refs.~\cite{RidingerDavidson07,RidingerWeiss09,ClearyEtAl10}). Recent related investigations of the influence of the initial phase of the driving [replacing $\cos(\omega t)$ in the Hamiltonian~(\ref{eq:H2time}) by $\cos(\omega t +\phi)$] can be found in Refs.~\cite{CreffieldSols11,KudoMonteiro11,ArlinghausHolthaus11}.

In the following, the time-reversal is demonstrated by numerically solving the full, time-dependent Hamiltonian~(\ref{eq:H2time}) corresponding to the ideal time-reversal Hamiltonian~(\ref{eq:hdream})  using the Shampine-Gordon routine~\cite{ShampineGordon75}. 
\begin{figure}
\includegraphics[width=\linewidth]{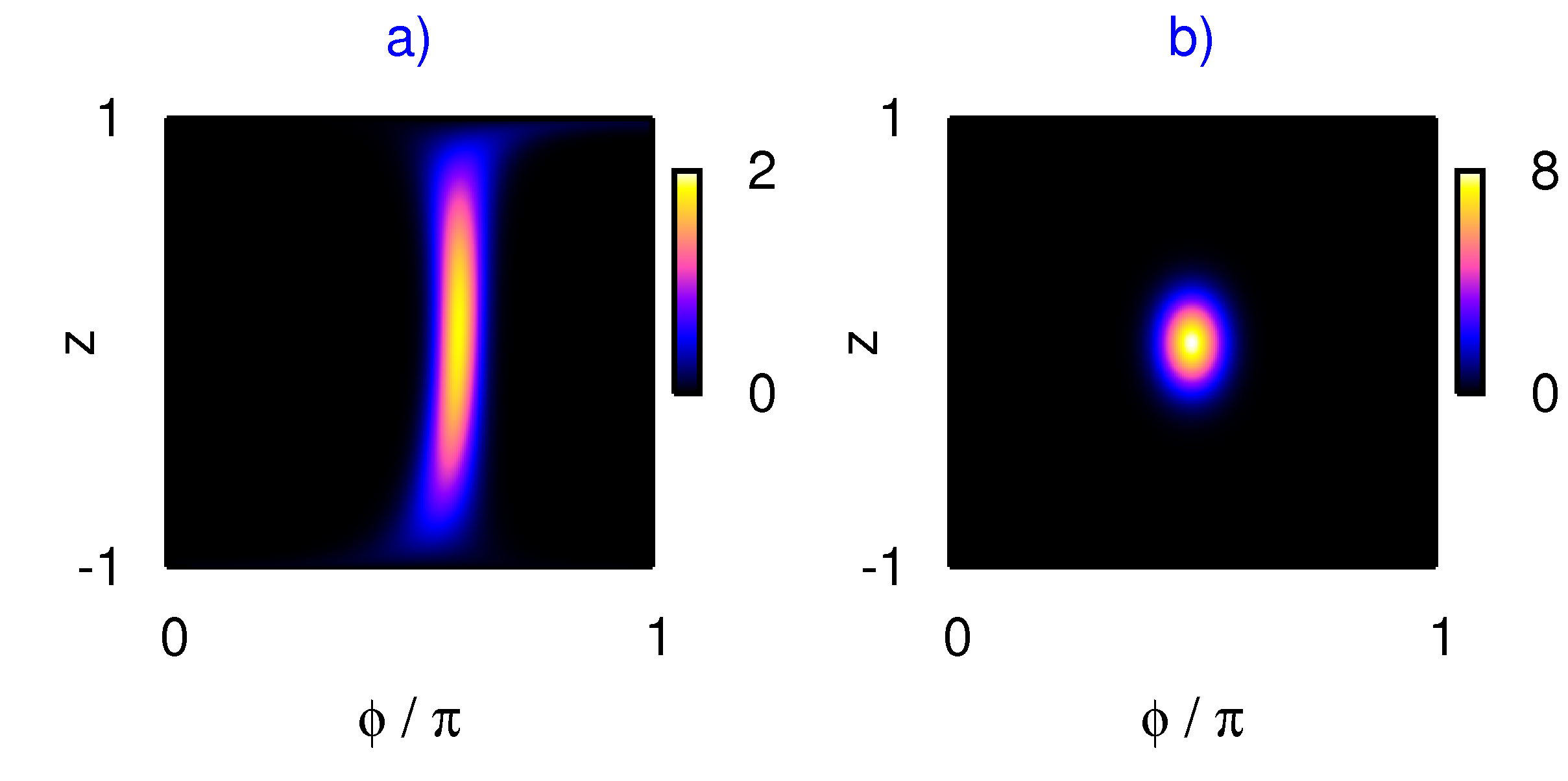}
\caption{\label{fig:welle}(Colour online) {Wave-function after apparent damping dynamics versus typical product state} \textbf{a)} Parameters as in Fig.~\ref{fig:damping} except for $K_0= 1.691732695$ and $\hbar\omega =32J$. The wave-function at $\tau=30\pi$ is displayed as a function of population imbalance~$z$ and phase~$\phi$ between the two wells [cf.\ Eq.~(\ref{eq:atomic})]. This quantum superposition can be characterised by $F_z\simeq 38.8$ and it could thus be used to improve interferometric measurements. \textbf{b)} If all 100 atoms occupy the same single particle state (here: $z=0$, $\phi=\pi/2$), the wave-function is much narrower and (as no product state satisfies $F_z> 1$) would not be interesting for interferometry.
}
\end{figure}
Contrary to time-reversal schemes on the level of the Gross-Pitaevskii equation~\cite{MartinEtAl08,TsukadaEtAl08}, here time-reversal is used to distinguish interesting quantum superpositions from statistical mixtures. Before implementing the time-reversal, Fig.~\ref{fig:welle} shows the wave-function for $N=100$ particles which were initially in one well. After several oscillations, the wave-function no longer is in a product state. Both the population imbalance and the phase can be measured experimentally~\cite{EsteveEtAl08}; in Fig.~\ref{fig:welle} the squared modulus of the scalar product with the atomic coherent states~\cite{MandelWolf95},
\begin{eqnarray}
\label{eq:atomic}
\left|\theta,\phi\right>_N&=& \sum_{n=0}^N 
{N \choose n}^{1/2}
\cos^{n}(\theta/2)
                 \sin^{N-n}(\theta/2)
                 \nonumber\\
                 &\times& e^{i(N-n)\phi}| n, N-n \rangle\,,
\end{eqnarray}
is plotted. The angle~$\theta$ corresponds to a population imbalance of
\begin{equation}
\frac{\langle z \rangle}2 = \cos(\theta)
\end{equation}

Ideally, it should be possible to show that the wave-function of Fig.~\ref{fig:welle}~(a) indeed is a quantum superposition by using the time-reversal of Eq.~(\ref{eq:hdream}) and than showing that
\begin{equation}
\langle z_{\rm end}\rangle  \equiv\langle z\rangle (2\tau_0)\;
\end{equation}
is one: There is only one many-particle wave-function for which this is the case. Furthermore, the unitary evolution of solutions of the Schr\"odinger equation guarantees that for two different solutions $|\psi_1(2\tau_0)\rangle=U(\tau_0,2\tau_0)|\psi_1(\tau_0)\rangle$ and  $|\psi_2(2\tau_0)\rangle=U(\tau_0,2\tau_0)|\psi_2(\tau_0)\rangle$, the scalar product would be the same at $\tau=\tau_0$ and at  $\tau=2\tau_0$ (as $U^{\dag}U=1$).

\begin{figure}
\includegraphics[width=\linewidth]{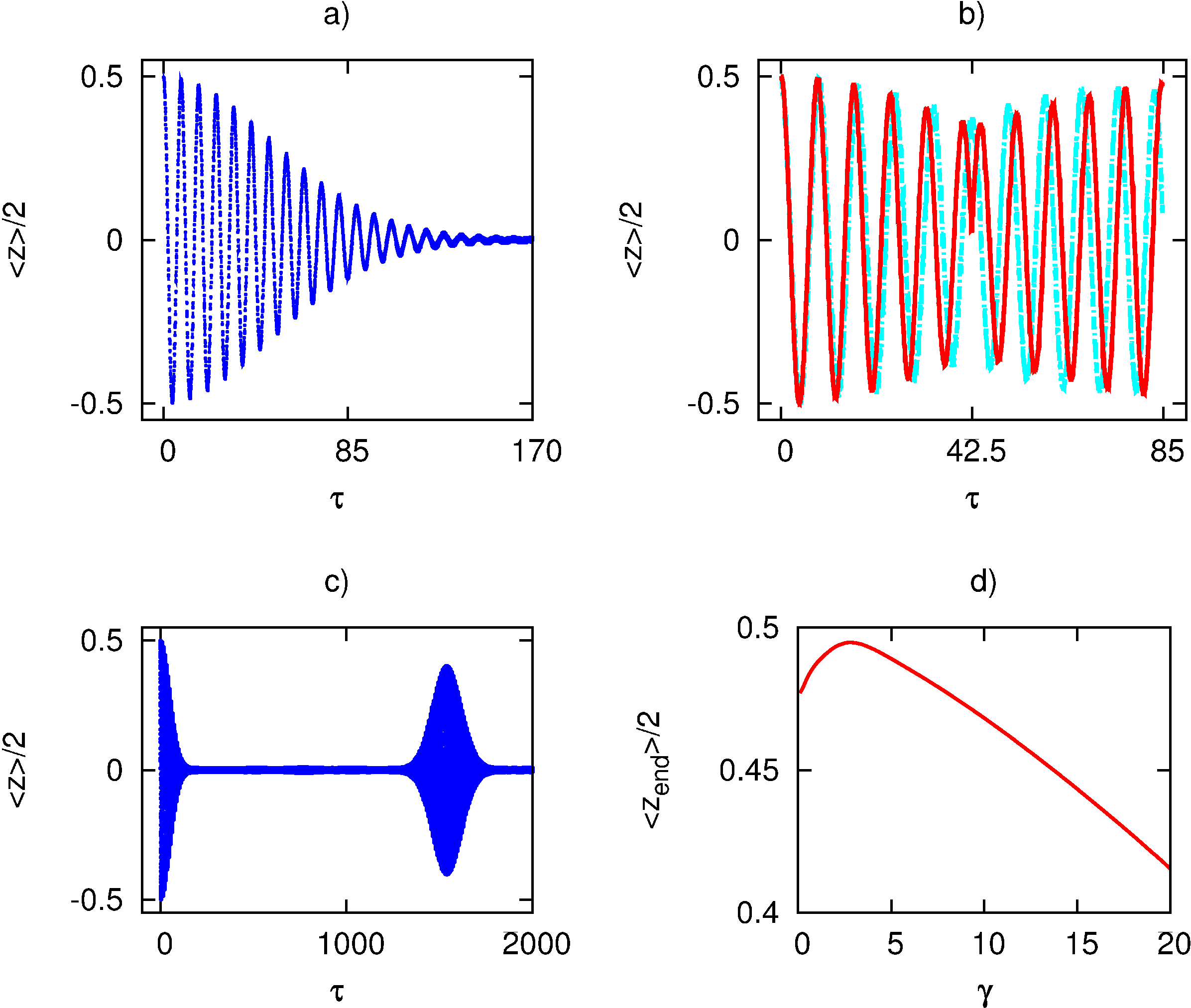}
\caption{\label{fig:allesrechts}(Colour online) {Time-reversal of the quantum dynamics of a small Bose-Einstein condensate in a periodically shaken double well [Eq.~(\ref{eq:H2time})].}  \textbf{a)} Population imbalance~(\ref{eq:popu}) as a function of time~$\tau=tJ/\hbar$ for the same parameters as in Fig.~\ref{fig:welle}~(b) 
\textbf{b)} 
  Red/dark solid line: all other parameters as in panel (a)  except for $\tau>\tau_0$: $K_0=3.831705970\,J$ and $U = -0.4J/N$; light blue/grey dash-dotted line: as red/dark line but $\hbar\omega =12J$; in both cases the revival of the initial state is visible near $\tau\approx 85$.   \textbf{c)} Population imbalance for the same situation as in panel (a) but for much longer time-scales. \textbf{d)} If the switching takes place continuously rather than instantaneously [Eq.~(\ref{eq:switch})], the revival of the initial state can still be observed [same parameters as for the red/dark curve in panel (b)] ($\gamma = 0$ corresponds to instantaneous switching).}
\end{figure}

\begin{figure}
\includegraphics[width=\linewidth]{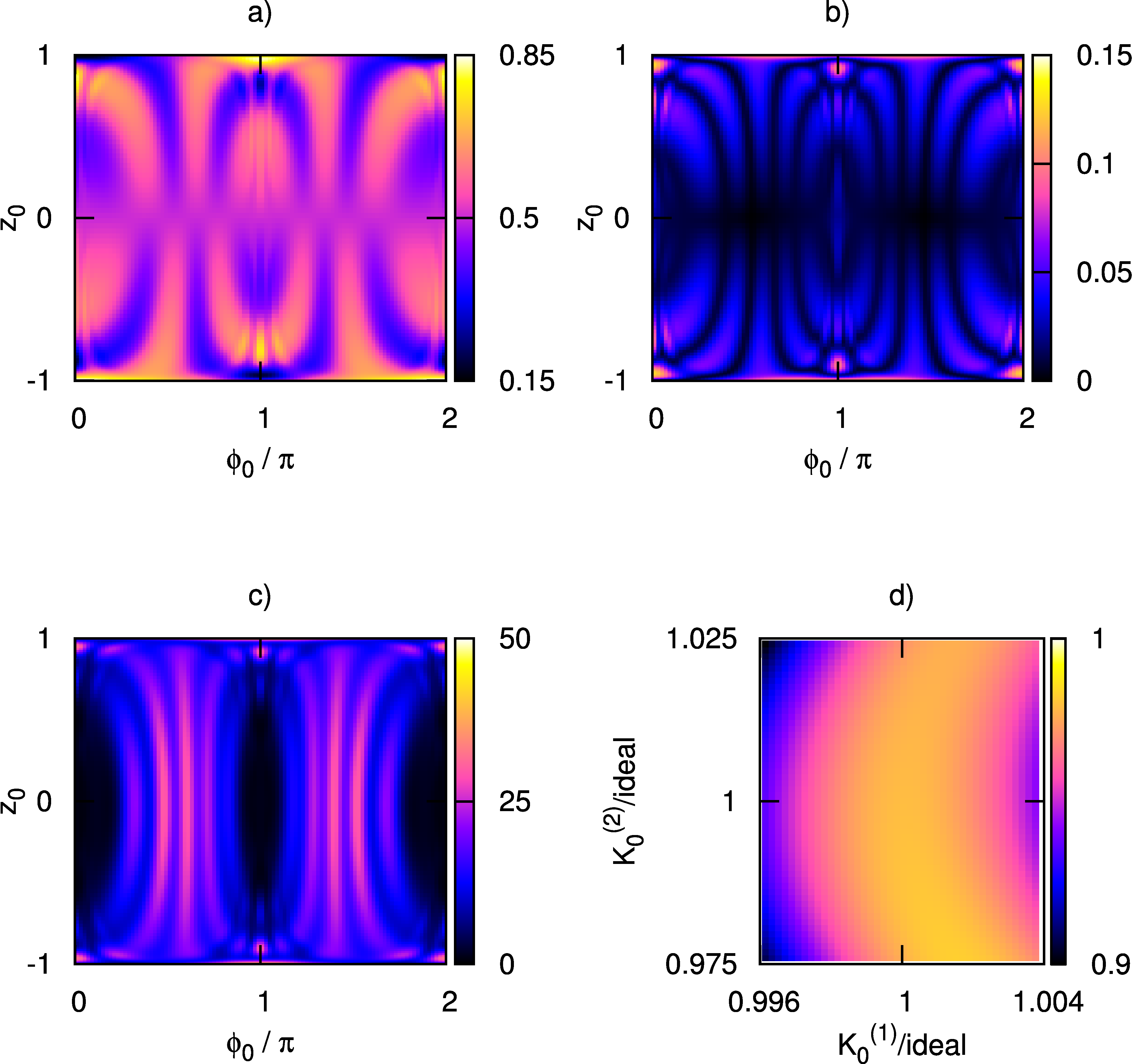}
\caption{\label{fig:karten}(Colour online) \textbf{a)} Two-dimensional projection of $\langle z_{\rm max}\rangle/2+0.5$ as a function of both the initial phase and the initial population imbalance [cf.\ Eq.~(\ref{eq:atomic}); $z_0 = \cos\left(\theta_0\right)$] For the numerics, the Hamiltonian leading to the red/dark curve in Fig.~\ref{fig:allesrechts}~(b) for $\tau>\tau_0$ was used. \textbf{b)} Two-dimensional projection of $\Delta z$ [Eq.~(\ref{eq:deltaz})] shows that $\langle z\rangle$ does not change dramatically on short time-scales. \textbf{c)} $F_z(\tau_0)$~(\ref{eq:Fz}) as a two-dimensional projection.  \textbf{d)} If the time-reversal scheme of Fig.~\ref{fig:allesrechts}~(b) for the red/dark curve is repeated with not ideal driving amplitudes, $\langle z_{\rm end}\rangle/2 +0.5$ (shown as a two-dimensional projection as a function of both driving amplitudes normalised by their ideal values - cf.\ Fig.~\ref{fig:sketch}) still lies well above the values shown in panel (a).
}
\end{figure}

However, the Hamiltonian~(\ref{eq:hdream}) is a high-frequency approximation and it has thus to be shown that this works for realistic driving frequencies (cf.\ Fig.~\ref{fig:allesrechts}). Furthermore, although there is only one wave-function at $\tau=\tau_0$ which exactly leads to the value $\langle z\rangle_{\rm end}=1$ at $\tau=2\tau_0$, other (less interesting) wave-functions might lead to values close to $\langle z\rangle_{\rm end}=1$ (cf.\ Fig.~\ref{fig:karten}). 

Figure~\ref{fig:allesrechts} shows that the time-reversal dynamics is indeed feasible. On time-scales for which there is not even a partial revival of the initial state characterised by $\langle z\rangle=1$, the proposed time-reversal dynamics leads to final values above $\langle z_{\rm end}\rangle/2=0.45$ (Fig.~\ref{fig:karten} shows that this is enough to show that the wave-function at $\tau=\tau_0$ was indeed a quantum superposition).
In order to show that the scheme does not rely on the switching to be truly instantaneous at $t=t_0$ [where $t_0$ is linked to $\tau_0$ via Eq.~(\ref{eq:tau})], the amplitude in Fig~\ref{fig:allesrechts}~(d) was switched according to
\begin{equation}
\label{eq:switch}
K_0(t)=K_0^{(1)}+\left(K_0^{(2)}-K_0^{(1)}\right)\frac{1+\tanh\left(\frac{\omega (t-t_0)}{\gamma}\right)}2;
\end{equation}
the switching between the two interaction values was chosen analogously (for instantaneous switching, the switching time can also slightly deviate from the ideal switching time, it just has to be close to the shaking maximum).

Figure~\ref{fig:karten} primarily investigates how close the value of $\langle z_{\rm end}\rangle$ has to be to one in order to show that the intermediate state shown in Fig.~\ref{fig:welle}~(a) really is a quantum superposition. Figure~\ref{fig:karten} investigates the time-dynamics of product states~(\ref{eq:atomic}) under the time-evolution which for the quantum superposition of Fig.~\ref{fig:welle}~(a) would lead to a revival of the initial state [Fig.~\ref{fig:allesrechts}~(b)].  Figure~\ref{fig:karten}~(a) shows that
\begin{equation}
\label{eq:zmax}
\langle z_{\rm max} \rangle  \equiv \left.\max\left\{\langle z\rangle (\tau)\right\}\right|_{0.99\tau_0\le\tau\le\tau_0}\;
\end{equation}
lies well below the values achieved in time-reversal [Fig.~\ref{fig:allesrechts}~(b)]. Furthermore, it does not change dramatically on short time-scales, as can be seen in Fig.~\ref{fig:karten}~(b) which uses
a $\langle z_{\rm min} \rangle$  which is analogously defined to Eq.~(\ref{eq:zmax}) to calculate
\begin{equation}
\label{eq:deltaz}
\Delta z \equiv \frac{\langle z_{\rm max} \rangle - \langle z_{\rm min} \rangle}2\,. 
\end{equation}
In addition to not approaching $\langle z\rangle=1$, many product states lead to very large fluctuations [Fig.~\ref{fig:karten}~(c)]; these fluctuations are particularly large if one compares them with the tiny values of $F_z(2\tau_0)\simeq 0.4$ for the red/dark curve in Fig.~\ref{fig:allesrechts}~(b). This offers an additional route to distinguish quantum superpositions as in Fig.~\ref{fig:welle}~(a) from statistical mixtures by carefully investigating how the product states~(\ref{eq:atomic}) with large contributions to Fig.~\ref{fig:welle}~(a) behave.
Figure~\ref{fig:karten}~(d) shows that the time-reversal scheme is feasible even if the driving amplitudes only approximately meet the ideal values [Fig.~\ref{fig:karten}~(d)].

To conclude, time-reversal via quasi-instantaneously changing the sign of the effective Hamiltonian is experimentally feasible for ultra-cold atoms in a periodically shaken  double well. The change of the sign of the Hamiltonian is achieved by changing both the driving amplitude and the sign of the interaction; a particularly useful initial state is the state with all particles in one well. The numeric investigations show that the revival of the initial state can be used to distinguish damping introduced via decoherence from the apparent damping related to a collapse phenomenon. Even if the revival of the initial state is not perfect, the scheme clearly distinguishes product states from quantum superpositions with potential interferometric applications.

\acknowledgements
I would like to thank S.A.\ Gardiner and M.\ Holthaus for their support and T.P.\ Billam, B.\ Gertjerenken,  E.\ Haller, C.\ Hoffmann, J.\ Hoppenau, A.\ Ridinger, T.~Sternke and S.~Trotzky for discussions.

%


\end{document}